\documentclass[pre, twocolumn, amsmath,superscriptaddress, floatfix, byrevtev]{revtex4-1}
\usepackage{times}
\usepackage{mathptmx}
\usepackage[dvips]{graphicx}

\usepackage{graphicx}% Include figure files
\usepackage{dcolumn}% Align table columns on decimal point
\usepackage{bm}% bold math
\usepackage{multirow}
\usepackage{color}
\usepackage{comment}

\begin{document}
\title{Super-ponderomotive electron acceleration in blowout plasma heated by multi-picosecond relativistic intensity laser pulse}
%\title{High energy electron acceleration by multi-ps relativistic laser pulse assisted by both self-generated electric field and magnetic field in expanding plasmas}
% Force line breaks with \\
%\title{ENERGY DISTRIBUTION OF FAST ELECTRONS GENERATED WITH RELATIVISTIC INTENSITY LASER DEPENDING ON PULSE DURATION}% Force line breaks with \\
%\thanks{This is a draft of a paper to be submitted to physical review letters.}%

\author{Sadaoki Kojima}
\email{skojima@laser.kuicr.kyoto-u.ac.jp}
\affiliation{Institute of Laser Engineering, Osaka University, 2-6 Yamada-Oka, Suita, Osaka, 565-0871 Japan.}
\author{Masayasu Hata}
\affiliation{Institute of Laser Engineering, Osaka University, 2-6 Yamada-Oka, Suita, Osaka, 565-0871 Japan.}
\author{Natsumi Iwata}
\affiliation{Institute of Laser Engineering, Osaka University, 2-6 Yamada-Oka, Suita, Osaka, 565-0871 Japan.}
\author{Yasunobu Arikawa}
\affiliation{Institute of Laser Engineering, Osaka University, 2-6 Yamada-Oka, Suita, Osaka, 565-0871 Japan.}
\author{Alessio Morace}
\affiliation{Institute of Laser Engineering, Osaka University, 2-6 Yamada-Oka, Suita, Osaka, 565-0871 Japan.}
\author{Shouhei Sakata}
\affiliation{Institute of Laser Engineering, Osaka University, 2-6 Yamada-Oka, Suita, Osaka, 565-0871 Japan.}
\author{Seungho Lee}
\affiliation{Institute of Laser Engineering, Osaka University, 2-6 Yamada-Oka, Suita, Osaka, 565-0871 Japan.}
\author{Kazuki Matsuo}
\affiliation{Institute of Laser Engineering, Osaka University, 2-6 Yamada-Oka, Suita, Osaka, 565-0871 Japan.}
\author{King Fai Farley Law}
\affiliation{Institute of Laser Engineering, Osaka University, 2-6 Yamada-Oka, Suita, Osaka, 565-0871 Japan.}
\author{Hiroki Morita}
\affiliation{Institute of Laser Engineering, Osaka University, 2-6 Yamada-Oka, Suita, Osaka, 565-0871 Japan.}
\author{Yugo Ochiai}
\affiliation{Institute of Laser Engineering, Osaka University, 2-6 Yamada-Oka, Suita, Osaka, 565-0871 Japan.}
\author{Akifumi Yogo}
\affiliation{Institute of Laser Engineering, Osaka University, 2-6 Yamada-Oka, Suita, Osaka, 565-0871 Japan.}
\author{Hideo Nagatomo}
\affiliation{Institute of Laser Engineering, Osaka University, 2-6 Yamada-Oka, Suita, Osaka, 565-0871 Japan.}
\author{Tetsuo Ozaki}
\affiliation{National Institute for Fusion Science, National Institutes of Natural Sciences, 322-6 Oroshi, Toki, Gifu, 509-5292, Japan.}
\author{Tomoyuki Johzaki}
\affiliation{Department of Mechanical Systems Engineering, Hiroshima University, Higashi-Hiroshima, Hiroshima, 739-8527, Japan.}
\author{Atsushi Sunahara}
\affiliation{Institute for Laser Technology, 1-8-4 Utsubo-honmachi, Nishi-ku Osaka, Osaka, 550-0004, Japan.}
\author{Hitoshi Sakagami}
\affiliation{National Institute for Fusion Science, National Institutes of Natural Sciences, 322-6 Oroshi, Toki, Gifu, 509-5292, Japan.}
\author{Zhe Zhang}
\affiliation{Beijing National Laboratory of Condensed Matter Physics, Institute of Physics, Chinese Academy of Sciences, Beijing 100190, China.}
\author{Shota Tosaki}
\affiliation{Institute of Laser Engineering, Osaka University, 2-6 Yamada-Oka, Suita, Osaka, 565-0871 Japan.}
\author{Yuki Abe}
\affiliation{Institute of Laser Engineering, Osaka University, 2-6 Yamada-Oka, Suita, Osaka, 565-0871 Japan.}
\author{Junji Kawanaka}
\affiliation{Institute of Laser Engineering, Osaka University, 2-6 Yamada-Oka, Suita, Osaka, 565-0871 Japan.}
\author{Shigeki Tokita}
\affiliation{Institute of Laser Engineering, Osaka University, 2-6 Yamada-Oka, Suita, Osaka, 565-0871 Japan.}
\author{Mitsuo Nakai}
\affiliation{Institute of Laser Engineering, Osaka University, 2-6 Yamada-Oka, Suita, Osaka, 565-0871 Japan.}
\author{Hiroaki Nishimura}
\affiliation{Institute of Laser Engineering, Osaka University, 2-6 Yamada-Oka, Suita, Osaka, 565-0871 Japan.}
\author{Hiroyuki Shiraga}
\affiliation{Institute of Laser Engineering, Osaka University, 2-6 Yamada-Oka, Suita, Osaka, 565-0871 Japan.}
\author{Hiroshi Azechi}
\affiliation{Institute of Laser Engineering, Osaka University, 2-6 Yamada-Oka, Suita, Osaka, 565-0871 Japan.}
\author{Yasuhiko Sentoku}
\affiliation{Institute of Laser Engineering, Osaka University, 2-6 Yamada-Oka, Suita, Osaka, 565-0871 Japan.}
\author{Shinsuke Fujioka}
\email{sfujioka@ile.osaka-u.ac.jp}
\affiliation{Institute of Laser Engineering, Osaka University, 2-6 Yamada-Oka, Suita, Osaka, 565-0871 Japan.}

\begin{abstract}
The dependence of the mean kinetic energy of laser-accelerated electrons on the laser intensity, so-called ponderomotive scaling, was derived theoretically with consideration of the motion of a single electron in oscillating laser fields.
This scaling explains well the experimental results obtained with high-intensity pulses and durations shorter than a picosecond; however, this scaling is no longer applicable to the multi-picosecond (multi-ps) facility experiments.
Here, we experimentally clarified the generation of the super-ponderomotive-relativistic electrons (SP-REs) through multi-ps relativistic laser-plasma interactions using prepulse-free LFEX laser pulses that were realized using a plasma mirror (PM). The SP-REs are produced with direct laser acceleration assisted by the self-generated quasi-static electric field and with loop-injected direct acceleration by the self-generated quasi-static magnetic field, which grow in a blowout plasma heated by a multi-ps laser pulse.
Finally, we theoretically derive the threshold pulse duration to boost the acceleration of REs, which provides an important insight into the determination of laser pulse duration at kilojoule- petawatt laser facilities.

%[Note: This abstract is almost twice the word length of that recommended (The first part highlighted in yellow could be deleted to help reduce the word count, although the word count would still exceed the recommendation). From Instructions to authors: The abstract is typically 150 words and is unreferenced; it contains a brief account of the background and rationale of the work, followed by a statement of the main conclusions introduced by the phrase “Here we show” or some equivalent.]

%\begin{description}
%\end{description}
\end{abstract}

%\tableofcontents
\maketitle

\section{Introduction}
%[Note: No heading please: An introduction (without heading) of up to 500 words of referenced text expands on the background of the work (some overlap with the summary is acceptable), followed by a concise, focused account of the findings, ending with one or two short paragraphs of discussion.] 

When a high-intensity laser pulse is irradiated on a material, its surface is instantaneously ionized, and the electrons in the ionized material, i.e., plasma, are then accelerated close to the speed of light by the ponderomotive force of the laser light.
These energetic electrons are often called relativistic electrons (REs).
The energy distribution of REs is approximated by a Maxwell-Boltzmann distribution function with slope temperature $T_\textrm{RE}$ as $dN/dE \propto \exp \left( -E/T_\textrm{RE} \right)$ where $N$ and $E$ denote the number and energy, respectively. 
The scaling laws of $T_\textrm{RE}$ on laser intensity have been investigated experimentally \cite{Malka1996, Beg1997, Tanimoto2009}, theoretically, and computationally \cite{Wilks1992a, Haines2009, Kluge2011, Pukhov1999}.
These scaling laws are useful to determine laser parameters for high-intensity short pulse laser experiments and to design applications. 
The effect of pulse duration on $T_\textrm{RE}$ is not considered explicitly in the reported scaling laws; however, recent computational and theoretical studies \cite{Kemp2012, Sorokovikova2016} have revealed that $T_\textrm{RE}$ generated by multi-picosecond (multi-ps) laser pulses could be several times higher than that predicted by the reported scaling laws.
With the development of kilojoule-class high-power lasers such as LFEX \cite{Miyanaga2006}, NIF-ARC \cite{Crane2010}, LMJ-PETAL \cite{Batani2014}, and OMEGA-EP \cite{Maywar2008}, it has become possible to irradiate relativistic laser pulses continuously over multi-ps.

In this study, we have clarified the generation of super-ponderomotive RE (SP-RE) in multi-ps laser-plasma interaction using ultra-high-contrast LFEX laser pulses realized using a plasma mirror (PM). 
The slope temperature of REs was increased more than twice by extending the laser pulse duration from 1.2 ps to 4.0 ps.
The following two acceleration mechanisms were identified as essential for the generation of SP-REs in multi-ps laser-plasma interaction with the help of particle-in-cell (PIC) simulations.

One mechanism is the generation of SP-REs by the combination of a laser field and a quasi-static electric field reported by Sorokovikova \textit{et al.} \cite{Sorokovikova2016}.
In a laser-heated plasma, a quasi-static electric field is generated spontaneously by charge separation at the forward edge of the plasma expansion, and the direction of this field is generally parallel to the direction of laser propagation.
Such a quasi-static electric field is able to push electrons along the laser propagation direction; therefore, electrons can stay in the acceleration phase longer than that without a quasi-static electric field, i.e., electrons undergo higher energy gain.

The other mechanism is multiple electron injection in the region where the laser field and quasi-static electric field coexist due to the cyclotron motion of REs in a self-generated quasi-static azimuthal magnetic field \cite{Krygier2014, Nakamura2010}.
This distinctive injection mechanism is referred to as loop-injected direct acceleration (LIDA).
A tens of megagauss (MG) quasi-static magnetic field also develops within the expanding plasma in multi-ps laser-plasma interaction and LIDA plays a significant role in the generation of SP-REs in multi-ps laser-plasma interaction.
The LIDA is triggered by the transition from the hole boring phase to the blowout phase in a laser-heated plasma.
Here, we obtained the equation of transition timing for arbitrary laser pulses.

\section{Experimental observation of super-ponderomotive electrons}
We have experimentally investigated the dependence of RE energy distributions on the pulse durations under conditions free from pre-plasma formation.
The experiment was conducted using the LFEX laser system at the Institute of Laser Engineering, Osaka University.
The LFEX laser consists of four beams, where the spot diameter of the spatially overlapped LFEX beams on a target was 70 $\mathrm{\mu m}$ of the full width at half maximum (FWHM), and 30\% of the laser energy was contained in this spot.
One LFEX beam delivered 300 J of 1.053 $\mu$m wavelength laser light with a 1.2 ps duration (FWHM), and the peak intensity of one beam was 2.5$\times$10$^{18}$ $\rm{W/cm^2}$.

It is well known that SP-REs can be accelerated in a long-scale-length pre-plasma; therefore, a PM \cite{Doumy2004} was implemented to realize the pre-plasma-free condition to exclude the other known mechanisms from this experiment.
The contrast ratio of the LFEX laser pulse was improved by two orders of magnitude through implementation of the PM \cite{Arikawa2016} down to 10$^{11}$ at 150 ps before the main pulse (as shown in Fig.\,\ref{fig:experimental_setup}(a)).
These clean intense laser pulses create the ideal situation where the REs are accelerated predominantly in the inherent plasma formed by the main laser pulse itself during the picosecond time range.
The density scale length of the preformed plasma was calculated to be 1.5 $\mu$m at 10 ps before the intensity peak from a 2D radiation hydrodynamics simulation with the PINOCO-2D code \cite{Nagatomo2007}.

These ``clean" pulses were focused on a 1 mm$^3$ gold cube. 
The thickness of the gold cube is also an important parameter to investigate RE acceleration by multi-ps laser-plasma interactions.
The REs generate a sheath electric field at the rear surface of the target.
This sheath field refluxes especially low energy REs and the refluxed REs are re-injected to the acceleration region.
This recirculation process also generates SP-REs, which was investigated by Yogo and Iwata \textit{et al.} \cite{Yogo2017, Iwata2017}.
One cycle of the recirculation process takes at least 6.7 ps in the 1 mm-thick gold cube, which is longer than the pulse durations (1.2 or 4.0 ps) in this experiment; therefore, the recirculation process can be eliminated from the SP-RE mechanisms in this study.

LFEX laser pulses can be stacked temporally with arbitrary delays between the beams, as shown in Fig.\,\ref{fig:experimental_setup}(b).
In this study, a single beam (case A: 1.2 ps FWHM pulse duration and peak intensity of $2.5\times10^{18}$ $\rm{W/cm^2}$) was used and two types of four-stacked beams (case B: 4.0 ps FWHM pulse envelope and peak intensity of $3.0\times10^{18}$ $\rm{W/cm^2}$, and case C: 1.2 ps FWHM pulse duration and peak intensity of $1.0\times10^{19}$ W/cm$^2$).
We emphasize here that the leading edge of the stacked pulse remains similar to that of the single beam. If the pulse duration is extended by adjusting the pulse compressor of the laser system, the leading edge would inevitably be modified into a more gradual shape.

The energy distribution of REs emanated from the target to the vacuum was measured with an electron energy analyzer located 20.9$^{\circ}$ from the incident axis of the LFEX laser.
Figure \ref{fig:experiment}(a) shows the experimental results of the time-integrated energy distribution. 
The slope temperatures were 0.65 MeV for case A (red circles) and 1.7 MeV for case B (green triangles).
The slope temperature for case B was more than twice that for case A, even though the peak intensities were very close.
The energy distributions of REs obtained for case B (green triangles) and case C (blue squires) were almost identical, even though the peak intensities were different by a factor of four.
These slope temperatures cannot be explained using the reported scaling laws, whereby the dependence of the slope temperature on the pulse duration is not considered.

\section{Two-dimensional (2D) PIC simulations with experimental conditions}
\subsection{Electron acceleration dynamics in multi-picosecond laser-plasma interaction}
The experimental results were compared with those computed using the 2D PIC simulation code (PICLS-2D \cite{Sentoku2008}).
Calculations were performed with temporal and spatial scales that were comparable to the experimental scales.
The gold cube was replaced with a 20 $\mu$m planar plasma with a peak density of 40$n_{c}$, where $n_{c}$ = 1.0$\times$10$^{21}$ cm$^{-3}$ is the critical electron density for 1.053 $\mu$m wavelength light.
The bulk plasma has an exponential density profile from 0.1 to 40$n_{c}$ and a scale length of 1 $\mu$m. 
Due to computational limitations, the ionization degree was fixed to be +40 in the PICLS-2D simulation, which was determined based on the result of a one-dimensional PICLS simulation with the dynamic ionization model of gold described by field ionization \cite{Kato1998} and a fast electron collisional ionization \cite{Lotz1970}. The ionization degree rose from +10 (given by the radiation hydrodynamic code PINOCO-2D) to around +40 for first several picoseconds.
 In the 2D-PIC simulation with dynamic ionization, it was reported that ionizing defocusing counteracting laser filamentation and self-focusing occurs.% [Provide reference.]
 However, it does not significantly affect the short-scale-length pre-plasma in the order of the laser wavelength.

\begin{figure*}
	\begin{center}
 	\includegraphics*[width=160mm]{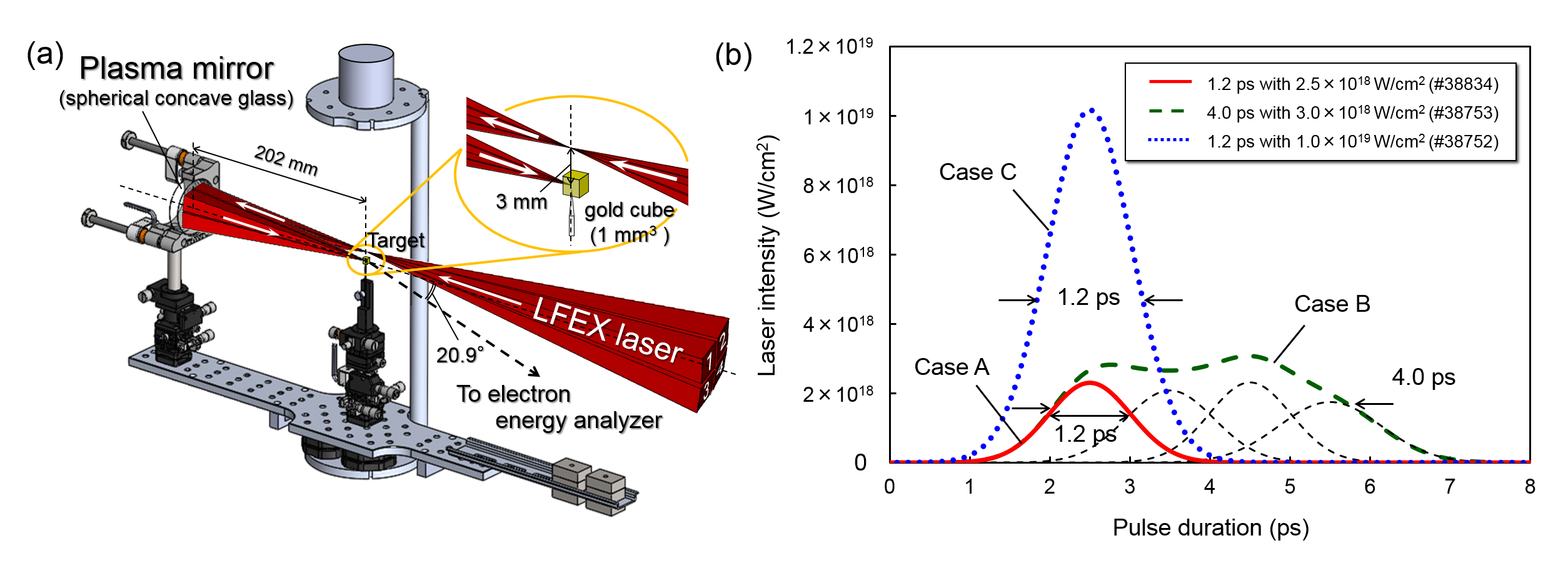}
	\end{center}
 \caption{(Color online) (a) Experimental setup. The geometrical positions of the target, PM and the diagnostics instruments, and the ray trace are illustrated. (b) Temporal intensity profiles of LFEX laser pulses. Pulses temporally stacked to generate various pulse shapes.
} \label{fig:experimental_setup}
\end{figure*}

\begin{figure*}
	\begin{center}
 	\includegraphics*[width=160mm]{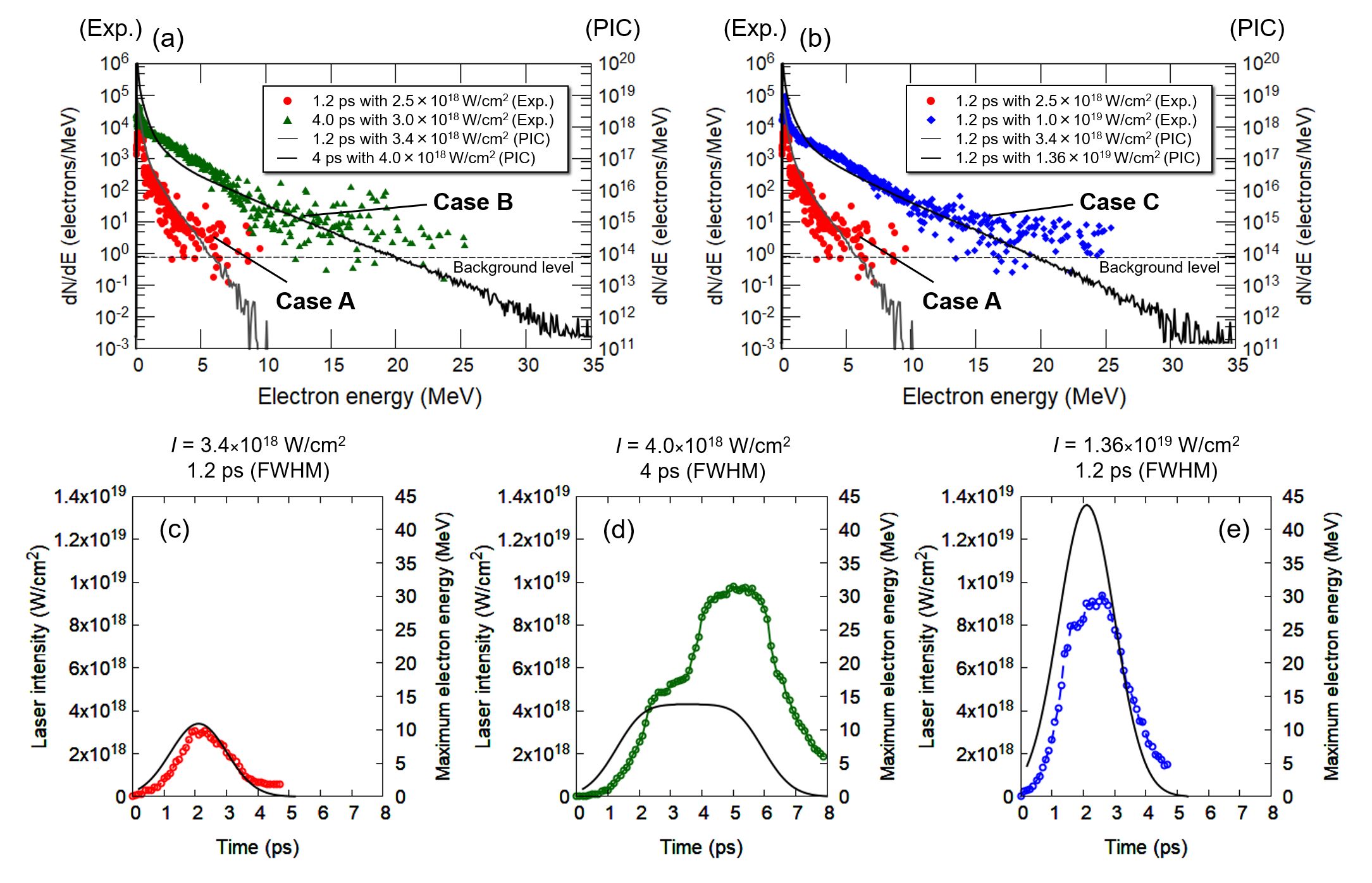}
	\end{center}
 \caption{(Color online)  RE energy distributions measured experimentally and computationally by changing the intensity and duration of laser pulses.
(a) Comparison of experimental data for cases A (red circles) and B (green triangles), and computational data for cases A (grey line) and B (black line).
(b) Comparison of experimental data for cases A (red circles) and C (blue squares), and computational data for cases A (grey line) and C (black line).
Comparison between laser pulse shapes (red lines) and temporal evolution of the maximum energy of REs (lines between circles) for cases (c) A, (d) B, and (e) C. 
} \label{fig:experiment}
\end{figure*}

The slope temperatures of the REs in the simulation were 0.7, 2.0, and 2.0 MeV for cases A, B, and C, respectively.
Thus, the PIC simulation reproduces well the experimentally observed dependence of the slope temperature on the laser intensity and pulse duration \cite{Kojima2016a}, as shown in Fig.\,\ref{fig:experiment}.

Figures \ref{fig:experiment} (c)--(e) show a comparison of the pulse shapes (red lines) and the temporal evolution of maximum energy of REs (blue lines between circles) for cases A, B, and C.
The temporal evolution of the maximum energy of the REs is similar to the laser pulse shapes for the cases of 1.2 ps pulse duration (cases A and C).
In contrast, the situation for the 4.0 ps pulse duration (case B) is completely different.
For case B, the maximum energy increases, even after the timing when the laser intensity reaches the plateau at 2.0 ps.
The most energetic REs were produced near the end of the intensity plateau (5.5 ps).  
The time-integrated energy distributions of the REs for cases B and C seem to be identical; however, the temporal behavior of RE acceleration in case B is completely different from that in case C.

\begin{figure*}
 	\begin{center}
 	\includegraphics[width=160mm]{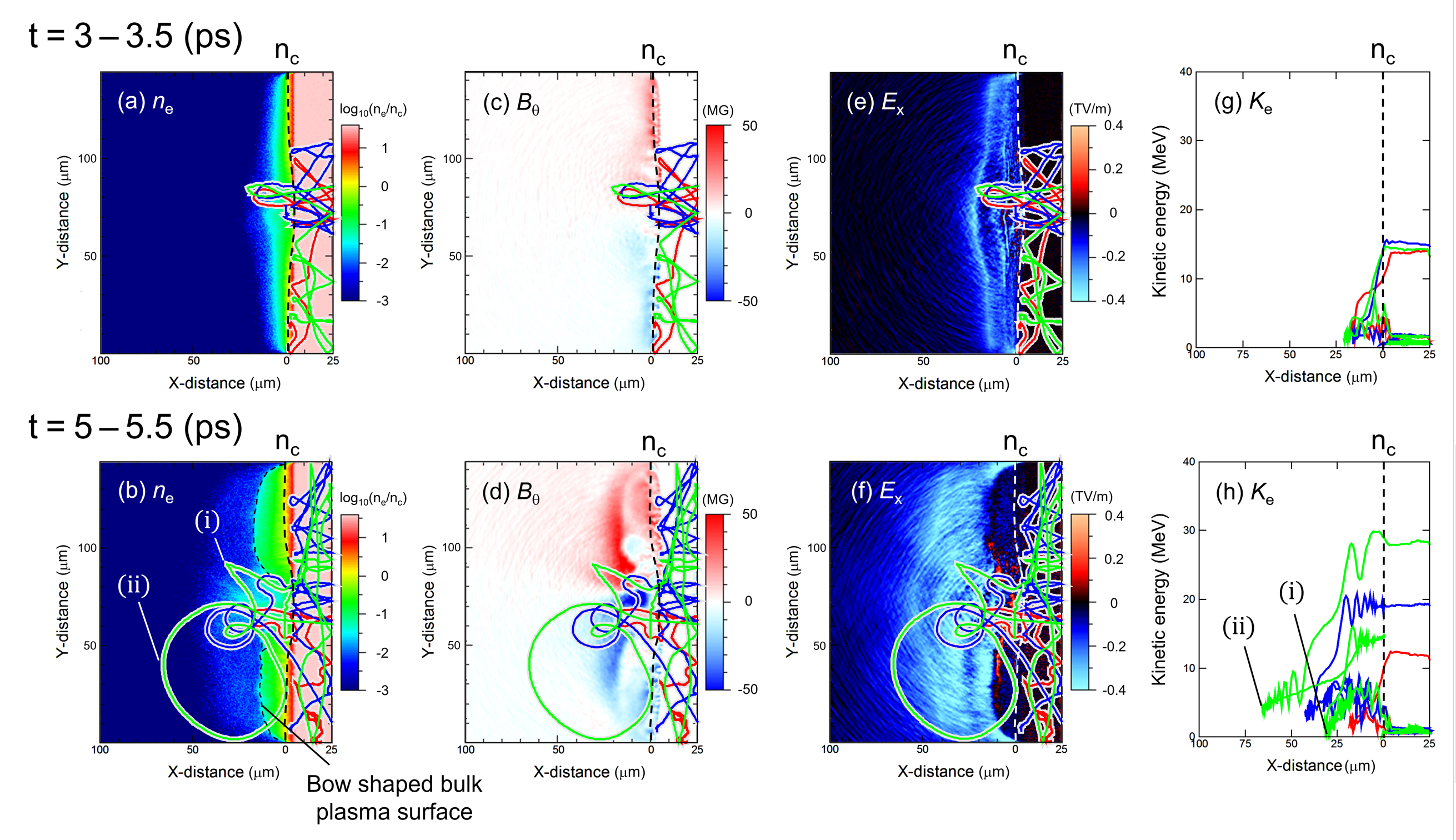}
 	\end{center}
 \caption{Three examples of RE trajectories at two different periods ($t$ = 3.0--3.5 and 5.0--5.5 ps) overlaid on the electron densities [(a) and (b)], self-generated azimuthal magnetic fields [(c) and (d)], and self-generated electric fields [(e) and (f)]. 
The electron density maps [(a) and (b)] are colored using the lookup table of electron density logarithm normalized according to the critical density ($n_\textrm{cr}$).
(g,h) Kinetic energies of REs along the longitudinal position for the two different periods. 
\\} \label{fig:track}
\end{figure*}

Figures \ref{fig:track}(a)--(f) show three selected RE trajectories at two different periods ($t$ = 3.0--3.5 and 5.0--5.5 ps) overlaid on the electron densities [Figs.\,\ref{fig:track}(a) and (b)], self-generated azimuthal magnetic fields [Figs.\,\ref{fig:track}(c) and (d)], and self-generated electric fields [Figs.\,\ref{fig:track}(e) and (f)]. 
Figures \ref{fig:track}(a) and (b) are colored using the lookup table of electron density logarithm normalized with the critical density ($n_\textrm{c}$).
When a high-intensity laser is irradiated on a target, quasi-static electric and magnetic fields are spontaneously generated on the target surface.
The quasi-static term indicates that the time variation of the fields is slower than that of the laser field.
The electric field is formed with plasma expansion and its direction is perpendicular to the target. The magnetic field is in the azimuthal direction of the laser axis.
These self-generated electric and magnetic fields assist RE acceleration as discussed below.

In the earlier period (the top panels of Fig.~\ref{fig:track}), the REs move around the near-critical density region.
The energetic RE source is initially accelerated to 3--4 MeV by the reflected laser field in the near-critical density region.
3--4 MeV is close to the kinetic energy (3.5 MeV) of a RE obtained by the ponderomotive force from the reflected laser field ($a_0 = 1.7$ and $I = 4.0\times10^{18} \mathrm{W/cm^2}$) without absorption of the incident laser field.
The electron travels outwardly (the opposite direction of laser propagation) through the magnetic and electric fields that are generated by the Biermann battery effect and charge separation.
In this period, the self-generated magnetic field strength is not sufficient to change the RE motion.
The self-generated electric field decelerates the outwardly moving RE, and the RE eventually stops and is then accelerated again inwardly.
The effect of the quasi-static electric field not only directly imparts additional energy to the electrons but also reduces the dephasing rate of the RE from the acceleration phase of the laser field  \cite{Sorokovikova2016, Robinson2013, Robinson2015, Paradkar2012a, Arefiev2016c, Arefiev2012, Kemp2009}. The RE continues to ride on the acceleration phase, whereby the RE gains energy from the laser field.
i.e., the RE obtains more energy when it is accelerated by the incident laser field.
In this simulation, the RE is accelerated up to 15 MeV by the combination of the quasi-static electric field and the laser field, as shown in Fig.\,\ref{fig:track}(g).

In the later period (bottom panels of Fig.\,\ref{fig:track}), the self-generated magnetic field is sufficiently strong that some of the REs (blue and green trajectories) are reflected outwardly by the $\bm{v} \times \bm{B}$ force and they are re-injected to the region where both the self-generated electric field and laser field coexist (Fig.\,\ref{fig:track}(b), loop(ii)).
In loop (ii), the turning point of the RE is farther from the near-critical density region than that in loop (i) because the RE receives more kinetic energy in loop (i).
After loop (ii), the kinetic energy of the REs reaches beyond 15 MeV, as shown in Fig.\,\ref{fig:track}(h).
This re-injection mechanism is the LIDA \cite{Krygier2014}. 

\begin{figure*}
	\begin{center}
	\includegraphics*[width=160mm]{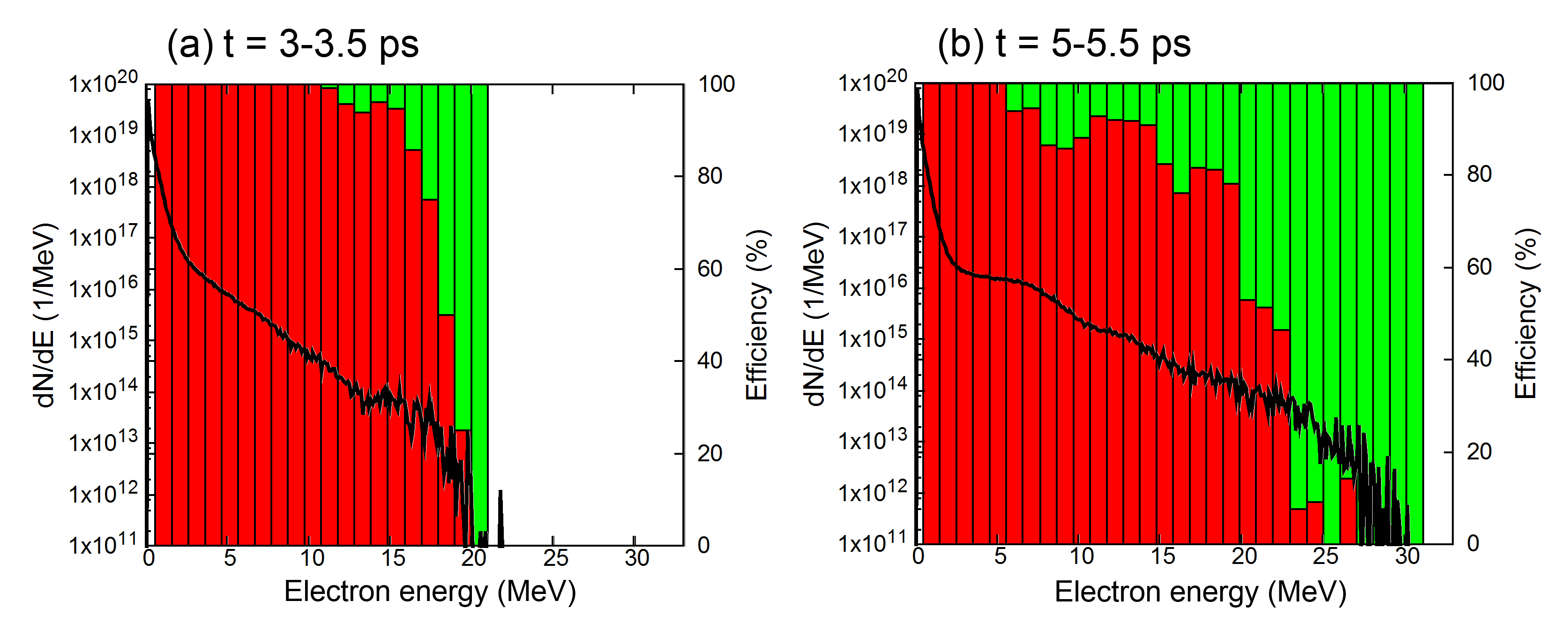}
	\end{center}
 \caption{(a,b) Energy distributions (solid lines) of REs accelerated in the two periods (3.0--3.5 and 5.0--5.5 ps).
The histograms show the ratio of the RE numbers between the two groups, where one group (red bars) consists of REs that experienced single loop-injection and another group (green bars) consists of REs that experience multiple loop-injections due to LIDA. 
A correlation between multiple loop-injections and energetic electron generation is clearly evident.
\\} \label{fig:histogram}
\end{figure*}

The solid lines in Figs.\,\ref{fig:histogram}(a) and (b) show energy distributions of REs accelerated in the two periods.
The histograms show the ratio of the RE numbers between the two groups: one group (red bars) consists of REs that experienced single loop-injection and the other (green bars) consists of REs that experienced multiple loop-injection. 
The correlation between multiple loop-injections and energetic RE generation is clearly evident; namely, the highest energy component of REs in Fig.\,\ref{fig:histogram}(b) above 20 MeV is generated predominantly by multiple loop-injection.

\subsection{Generation of a giant quasi-static magnetic field during multi-ps laser-plasma interaction}
The PIC simulation shows that the quasi-static magnetic field is generated by three different mechanisms in case B, which are dependent on the time during the multi-ps laser-plasma interaction.

\begin{figure*}
 \begin{center}
 \includegraphics*[width=160mm]{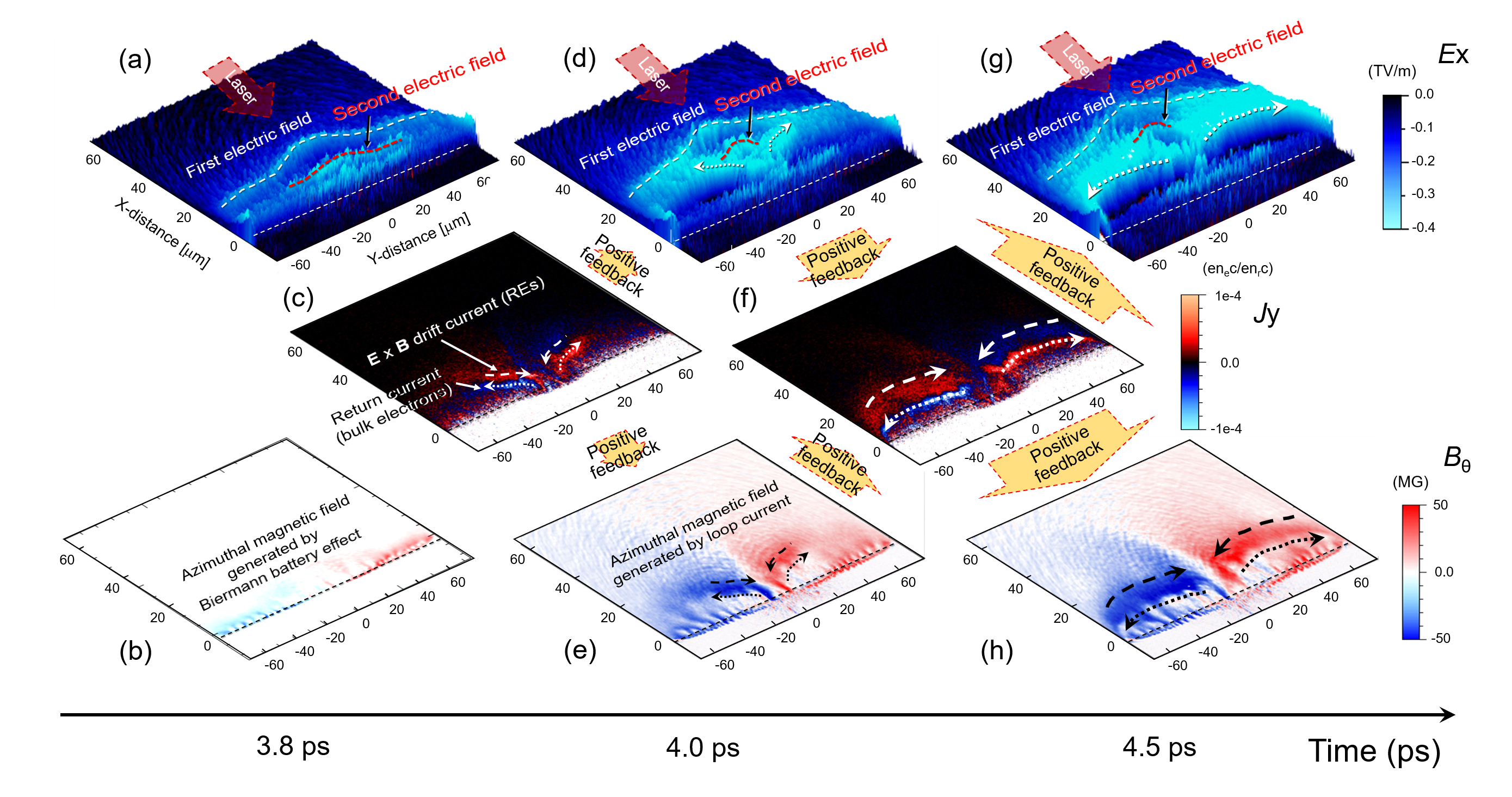}
 \end{center}
\vspace*{-0.5cm} 
 \caption{Spatial maps of longitudinal electric field $E_x$ [((a), (d), and (g)], transverse current density $J_y$ [(c) and (f)], and azimuthal magnetic field $B_\theta$ [(b), (e), and (h)] at 3.8, 4.0, and 4.5 ps. 
Loop current by the REs and return current rapidly enhance the strength of the electric and magnetic fields.
\\} \label{fig:Ex_Jy_Bz}
\end{figure*}

At the leading edge of the 4 ps flat-top pulse ($<$2 ps), the ponderomotive force of the incident laser pushes the relativistic critical density surface ($\gamma n_c$) into the overdense region, and the heated underdense plasma expands into the vacuum.
An electric field is generated at the outer boundary of the expanding plasma (which is referred to as the first electric field.).
An azimuthal magnetic field is generated in the overdense plasma due to the $\nabla n \times \nabla I$ effect \cite{Sudan1993, Mason1998, Wilks1992a, Tripathi1994}, where $n$ and $I$ are the plasma electron density and laser intensity, respectively.
When the laser intensity reaches the plateau at 2.0 ps, plasma evacuation by the laser field is eventually halted by the charge separation due to depletion of the local electron density. 
The $\nabla n \times \nabla I$ mechanism becomes relatively small, whereas the $\nabla T \times \nabla n$ (Biermann battery) effect \cite{Stamper1971, Borghesi1998, Max1978, Kolodner1979, Sandhu2002a, Schumaker2013a} becomes the dominant mechanism for generation of the magnetic field.
Here, $T$ is the plasma electron temperature. 
Along with a change of the generation mechanism, the generation region also moves from the overdense region to the underdense region.
The strongest magnetic field is generated at the edge of the laser spot in the underdense plasma (Fig.\,\ref{fig:Ex_Jy_Bz}(f)).
This magnetic field influences the motion of REs around the near-critical density region.
Some of the REs are moved transversely from the laser spot by the $\bm{E} \times \bm{B}$ drift.
The drift current heats the surface of the bulk plasma via the two-stream instability.
Enhancement of the energy transfer to the transverse direction due to the surface magnetic field is discussed in Refs.\,\cite{Jaanimagi1981,Paradkar2010a,Forslund1982}.
The electric field that contributes to the $\bm{E} \times \bm{B}$ drift is a weak electric field generated in a limited region near the critical density surface.
The heated bulk plasma begins to expand at the edge of the laser spot, while the expansion is suppressed at the inside of the laser spot by the laser ponderomotive pressure.
The heated bulk plasma surface, which has been flat so far, deforms into a bow shape (which is referred to as a bow-shaped bulk plasma surface).
The first electric field is carried out by the plasma expansion far away from the critical density surface and no longer contributes to the drift.

When the thermal pressure of the heated bulk plasma exceeds the ponderomotive pressure of the incident laser at 3.8 ps, the bulk plasma begins to expand at the inside of the laser spot, and the strong quasi-static electric field (the second electric field) is then generated at the near-critical density region.
Figure \ref{fig:Ex_Jy_Bz}(a) shows the electric fields in the longitudinal direction ($E_\mathrm{x}$) of the two regions.
The second electric field is generated at the expansion front of the heated bulk plasma at the inside of the laser spot.
The newly generated strong electric field contributes to the $\bm{E} \times \bm{B}$ drift by combination with the magnetic field (Fig.\,\ref{fig:Ex_Jy_Bz}(b)).
REs move along the bow-shaped bulk plasma surface by the $\bm{E} \times \bm{B}$ drift.
When the REs flow in the plasma, the return-current is driven to maintain current neutrality in the plasma.
Figure \ref{fig:Ex_Jy_Bz}(c) shows the RE drift current in the lower density region and the return-current flow in the higher density region.
The current loop produced by the spatial separation between the RE drift current and the return current generates a magnetic field along the outer edge of the bow-shaped bulk plasma surface (Fig.\,\ref{fig:Ex_Jy_Bz}(c)).
This third magnetic field (30--50 $\mathrm{MG}$) is stronger than the magnetic field generated by the $\nabla T \times \nabla n$ effect ($<$10 $\mathrm{MG}$). 
In the plasma region where RE current terminates, the electric field is enhanced by the inflow of electrons (Fig.\,\ref{fig:Ex_Jy_Bz}(d)).

The positive feedback between the growth of the fields and the field-driven drift current results in the rapid growth of the quasi-static electric and magnetic fields with time (Figs.\,\ref{fig:Ex_Jy_Bz}(c)--(h)).
The maximum energy of the REs increases from 3.5 ps until 5.5 ps, which corresponds to the timing of rapid growth of the self-generated fields.
The SP-RE are accelerated by a laser field under a quasi-static self-generated electric field.
In addition, when positive feedback starts, the strength of the self-generated magnetic field grows by several tens of MG approximately several picoseconds after the beginning of the laser-plasma interaction, and the strong magnetic field begins the LIDA.
Thus, SP-RE acceleration is not a process that gradually progresses with time but a process that proceeds in a threshold manner.
This has not been pointed out in previous studies on REs acceleration by multi-ps laser pulse. \cite{Sorokovikova2016,Kemp2012,Peebles2017,Yogo2017,Iwata2017}

\subsection{Transition timing to super-ponderomotive electron acceleration}
\begin{figure*}
 \begin{center}
 \includegraphics*[width=140mm]{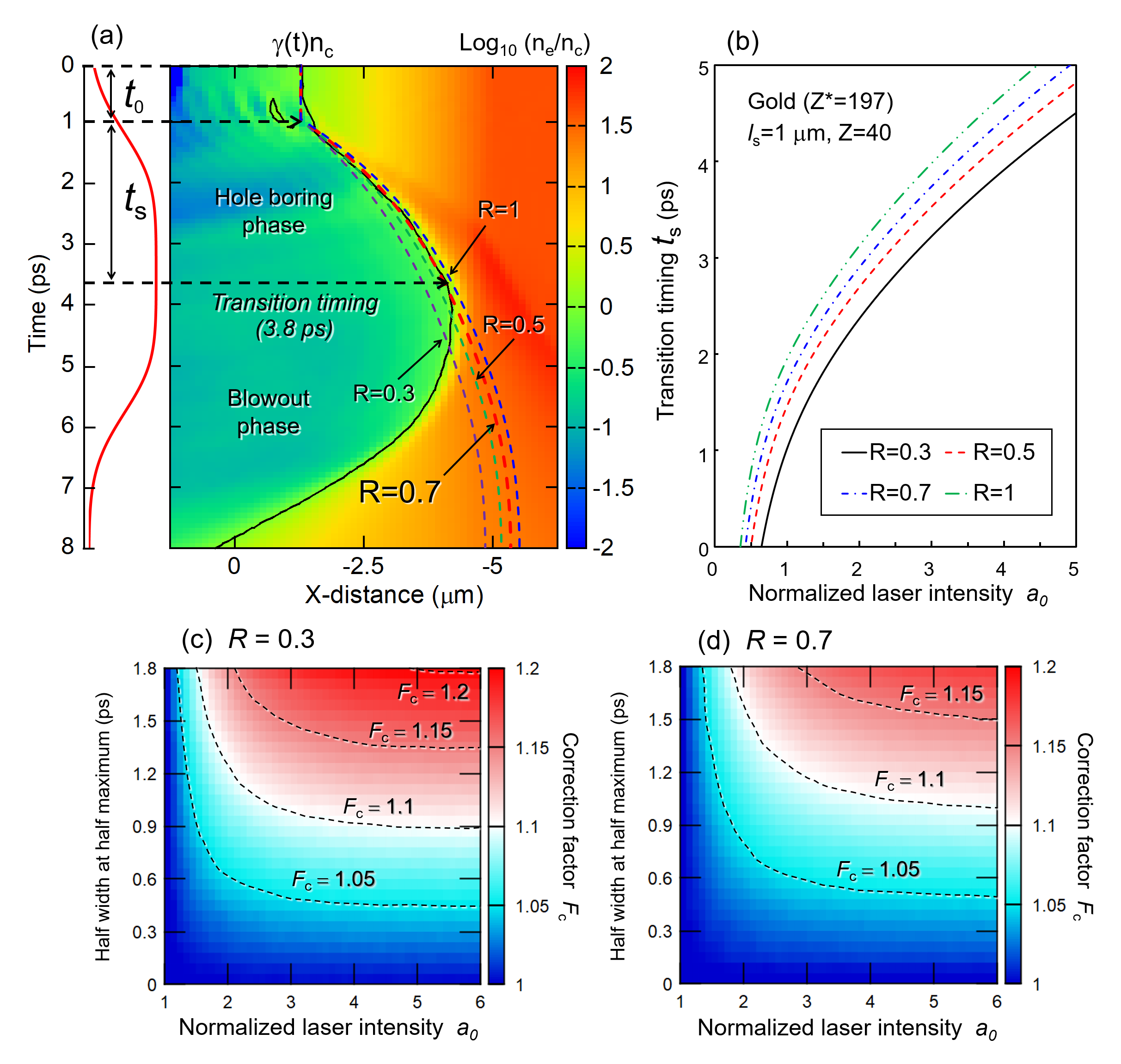}
 \end{center}
\vspace*{-0cm} 
 \caption{(a) Evolution of the initially exponential plasma profile during interaction with a laser pulse having a normalized laser intensity of $a_{0}=1.79$ ($I\lambda^{2}=4.0\times 10^{18}\,{\rm W\,\mu m^{2}/cm}^{2}$). The motion of the interface calculated from Eq.\,(\ref{eq: momentum conservation at the laser-plasma interaction surface_low}) (red dotted line) is in good agreement with the motion obtained by the PIC simulation until $t=$3.8\,ps (black solid line). (b) Transition timing for constant intensity temporal profile calculated from Eq.\,(\ref{eq: transition_timing_4}) for a gold target with a preformed plasma scale length of $\ell_{s}=1\,{\rm \mu m}$ in the charge state of $Z=40$. 
Correction factor of the pulse temporal profile $F_{c}$ for cases of (c) low reflectivity $R=0.3$ and (d) high reflectivity $R=0.7$.  
The normalized laser intensity $a_0$ or half width at half maximum (HWHM) of the Gaussian leading edge increases; therefore, a larger correction factor is required. 
} \label{fig:Plasma_compression}
\end{figure*}

The SP-RE acceleration is started when the plasma thermal pressure exceeds the laser ponderomotive pressure.
Figure \ref{fig:Plasma_compression}(a) shows the evolution of an initially exponential plasma profile during the interaction with a high-intensity laser pulse. The color map shows the electron density ($\log_{10}(n_e/n_c)$) and the red solid line shows the temporal intensity profile of the laser.
At $t=$3.8\,ps, the motion of the relativistic critical interface stops even though the laser pulse is still irradiated, and the state of the laser-plasma interaction transits from the hole boring phase to the blowout phase.

The position of the interface that interacts with the laser pulse having an arbitrary intensity temporal profile is obtained by integrating the velocity of the interface with respect to time \cite{Kemp2008},

\begin{widetext}
\begin{equation}
\begin{split}
x_i (t)= x_c(0) + 2l_s \ln \biggl[1+\frac{c}{2l_s} \sqrt{\frac{R \cos \theta}{(1+R)} \frac{Zm_e}{M_i }} \int_{t_0}^{t} \biggl( \frac{\gamma(t)^2 -1}{\gamma(t) }   \biggr)^{1/2} dt  \biggr]. \\
   \label{eq: momentum conservation at the laser-plasma interaction surface_low}
\end{split}
\end{equation}
\end{widetext}

Here, $I(t)/c=m_ec^2n_c a_0^2(t)/2$ is used and the variables are explained in the Methods section. $t_0$ is the time when the normalized laser amplitude $a_0$ reaches 1.
Note that the position of the interface $x_{c}$ should vary with time. 
However, here the initial position of the critical density, i.e., $x_{c}=x_{c}(0)=$constant, was substituted considering that the temporal profiles of realistic lasers increase from 0 to the peak intensity. 

The transition timing can be obtained by coupling Eq.\,\eqref{eq: momentum conservation at the laser-plasma interaction surface_low} with the hole boring limit density, which is derived from the momentum transfer equation for the stationary state of the interface \cite{iwata}:
\begin{equation}
\begin{split}
\frac{n_s}{n_c} = 8 \epsilon^2 a_0^2 \biggl[ \frac{1+R-(1-R)\beta_h^{-1}\alpha^{-1}}{2} \biggr],
   \label{eq: electron_number_density_plateau_hot_by_nc}
\end{split}
\end{equation} 
where $\epsilon$ is the polarization factor ($\epsilon$=1 and $\sqrt{2}$ for linear and circular polarization, respectively), the plasma is assumed to be composed of REs ($n_h$) and bulk electrons ($n_b$) as $n_e=n_h+n_b$, and the momentum flux of the bulk electron component is negligible compared to that of the RE component (i.e., $ n_e T_e c \beta_e \approx n_h T_h c\beta_h$). $\beta_h$ is the ratio of the drift velocity of REs ($v_h$) to the speed of light, $c$. $\alpha \equiv ir/2$ is the geometrical factor, where $r = 1$ for the non-relativistic Maxwell momentum distribution and $r = 2$ for the relativistic Maxwell (Maxwell-J\"uttner) momentum distribution. Here, $i =$ 1, 2, or 3 represents the dimension of the momentum distribution.
When a 1D relativistic Maxwell distribution $\alpha=1$ is assumed, the relativistic limit for the RE velocity $\beta_h=1$, and linear polarization $\epsilon=1$, Eq.\,(\ref{eq: electron_number_density_plateau_hot_by_nc})  reduces to $n_s/n_c = 8 R a_0^2$. 

By substituting $a_{0}=1.79$ and $R=0.7$, the electron density threshold $n_{s}$ for the experimental condition of case B in Fig.\,\ref{fig:Plasma_compression} is obtained as $n_{s}=17.9\,n_c$. This density is almost identical to the electron density threshold in which the plasma compression terminates in the PIC simulation. 
Substituting Eq.\,(\ref{eq: electron_number_density_plateau_hot_by_nc}) and the initial electron density profile ($n_e(x)=n_c \exp[(x-x_c)/l_s]$) into Eq.\,(\ref{eq: momentum conservation at the laser-plasma interaction surface_low}) yields 

\begin{widetext}
\begin{equation}
\begin{split}
8 \epsilon^2 a_0^2 \biggl[ \frac{1+R-(1-R)\beta_h^{-1}\alpha^{-1}}{2} \biggr] = \exp\Biggl\{ \frac{x_c +2l_s \ln \biggl[1+\frac{c}{2l_s}\sqrt{\frac{R\cos \theta}{(1+R)}\frac{Zm_e}{M_i}} \int_{t_0}^{t_s} \sqrt{\frac{\gamma(t)^2-1}{\gamma(t)}} dt \biggr]-x_c }{l_s}  \Biggr\},
  \label{eq: transition_timing_1}
\end{split}
\end{equation} 
\end{widetext}

where $a_0$ is the normalized laser intensity. When the laser intensity is constant in time, the transition timing is then obtained as 

\begin{widetext}
\begin{equation}
\begin{split}
t_s= F_c \biggl[ \frac{2l_s}{c}  \biggl\{ \sqrt{4 \epsilon^2 a_0^2 [ 1+R-(1-R)\beta_h^{-1}\alpha^{-1} ]}-1 \biggr\} \sqrt{\frac{(1+R)}{R\cos \theta}\frac{\alpha m_p Z^*}{Zm_e} \frac{\gamma}{\gamma^2-1}} \biggr] + t_0,
  \label{eq: transition_timing_4}
\end{split}
\end{equation} 
\end{widetext}

where $M_{i}=\alpha^* m_p Z^*$ represents the ion mass, $m_p$ is the proton mass, $Z^*$ is the ion charge number for the fully ionized state, and $\alpha^*=1$ for hydrogen and $\alpha^*=2$ for other species. 

The correction factor $F_c$ is added to take the laser pulse profile into account. 
For cases where the laser intensity is constant in time, $F_{c}=1$. 
The lines in Fig.\,\ref{fig:Plasma_compression}(b) show the transition timing calculated from Eq.\,(\ref{eq: transition_timing_4}) with $F_{c}=1$ for various reflectivities. 
Here, we assumed that a gold plasma ($Z^*=197$) with the preformed plasma scale length $l_s=$1 ${\rm \mu m}$ in the charge state of $Z=40$ interacts with a linearly polarized laser ($\epsilon=1$). The spot size of the LFEX laser is large; therefore, it is assumed that electron acceleration occurs one-dimensionally ($\alpha=1$).
As an approximate trend, when the normalized laser intensity $a_0$ or reflectivity $R$ increases, the transition timing is delayed. Low reflectivity reduces the effective laser intensity at the interface and reduces the hole boring limit density. 
When the normalized laser intensity is $a_0$=1.79 (intensity is $I\lambda^{2}=4.0\times 10^{18}\,{\rm W\,\mu m^{2}/cm}^{2}$), the transition timing is estimated to be $t_{s}=$2.8\,ps, which is in good agreement with the simulation result.

In an actual case, the laser intensity increases in time with the Gaussian profile, so that the transition timing is delayed compared with that obtained for $F_{c}=1$. 
The color maps in Figs.\,\ref{fig:Plasma_compression}(c) and (d) show the dependence of the correction factor $F_c$ on the normalized laser intensity $a_{0}$ and the half width at half maximum (HWHM) of the Gaussian leading edge for the low reflectivity case ($R$=0.3) and high-reflectivity case ($R$=0.7).
As the normalized laser intensity or HWHM increases, a larger correction factor is required. In the present range of $1 \le a_{0} \le 6$ and 0\,ps$\:\le$ HWHM $\le\:$1.8\,ps, the correction factor increases only approximately 1.2 times at the maximum; therefore, it is sufficient to use Eq.\,\eqref{eq: transition_timing_4} with $F_{c}=1$ for a rough estimation of the transition timing.

\section{Summary}
In summary, with the development of kilojoule-class high-power lasers, it has become possible to continuously irradiate relativistic laser pulses on matter over multi-ps.
Although electron acceleration using a conventional sub-ps laser pulse has been explained theoretically as the interaction of a single electron with a laser field, it is necessary to consider the collective effect of electrons when the pulse duration reaches the multi-ps range. 
Energetic RE generation was experimentally clarified with an average energy far beyond the ponderomotive scaling using the prepulse-free LFEX laser. 
During the multi-ps interaction, a quasi-static electric field is generated by plasma expansion. In addition, a quasi-static magnetic field is gradually generated due to three different mechanisms of the $\nabla n \times \nabla I$ effect, the $\nabla T \times \nabla n$ effect, and a loop current driven by the ${\textbf E}\times{\textbf B}$ drift.
The third mechanism of the current loop rapidly amplifies the magnetic field strength by the positive feedback between the electric and magnetic fields and the field-driven drift current.
Under the quasi-static electric field, REs are accelerated efficiently above ponderomotive scaling by the laser field because the dephasing rate of the REs from the laser field is reduced.
Furthermore, when the quasi-static magnetic field becomes sufficiently strong to reflect REs back to the laser-plasma interaction region, the reflected REs gain further additional energy.
The boosting timing of electron acceleration by the LIDA mechanism is related to the transition timing of the laser-plasma interaction state from the hole boring phase to the blowout phase.
The equation for transition timing can be derived from the equations for the motion of a relativistic critical density interface and the equations for the electron density where the hole boring terminates.
In this study, the equation for transition timing was extended to a laser pulse with an arbitrary intensity temporal profile. The theoretical result was then compared with the result of PIC simulation. 
The mechanism for the generation of SP-REs in the multi-ps laser-plasma interaction reported here is useful for various applications.
For instance, Yogo \textit{et al.} reported that the maximum proton energy is enhanced more than twice by extending the pulse duration to the multi-ps regime, due to the electron temperature evolution beyond the ponderomotive energy in the over picoseconds interaction. 
The acceleration mechanism of REs investigated here is also important for fast-ignition inertial confinement fusion and laboratory astrophysics using high-intensity multi-ps laser pulses.

\begin{acknowledgments}
The authors thank the technical support staff of the Institute of Laser Engineering (ILE) at Osaka University and those of the Plasma Simulator at the National Institute for Fusion Science (NIFS) for assistance with laser operation, target fabrication, plasma diagnostics, and computer simulations.
We also acknowledge A. Sagisaka, K. Ogura, A. S. Pirozhkov, M. Nishikino, and K. Kondo of the Kansai Photon Science Institute, National Institutes for Quantum and Radiological Science and Technology for valuable discussions on intensity contrast improvement using the PM.
This work was supported by the Collaboration Research Program between the NIFS and ILE at Osaka University, the ILE Collaboration Research Program, and by the Japanese Ministry of Education, Culture, Sports, Science and Technology (MEXT) through Grants-in-Aid for Scientific Research (Nos. 24684044, 24686103, 70724326, 15K17798, 25630419, 16K13918, and 16H02245), the Bilateral Program for Supporting International Joint Research of the Japan Society for the Promotion of Science (JSPS), and Grants-in-Aid for Fellows from JSPS (Nos. 14J06592, 17J07212 and 15J00850).
\end{acknowledgments}

\section*{Author contributions}
S. K. and S. F. are the principal investigators who proposed and organized the experiment. 
M. H. performed the PIC simulations in the collaboration with T. J. and H. S..
N. I., S. K., M. H. and Y. S. developed the theoretical model. 
Y. A. and A. M. designed and constructed the large size plasma mirror.
H. M. and Y. O. carried out the theoretical analysis.
H. N. and A. S. performed the radiation hydrodynamic simulations.
K. M., S. S., S. L., K. F. F. L. and Y. A. measured electron energy distribution of high energy component with some help from T. O..
S. T. measured electron energy distribution of low energy component with some help from Z. Z. and A. Y..\,
S. T. and J. K. are in charge of LFEX laser facility development in ILE.
M. N., H. N., H. S. and H. A. supervised the project and provided overall guidance.
All authors participated in the discussions and contributed to the preparation of the manuscript. 

\section{Methods}
\subsection{PM implementation}
The LFEX parabola cannot be focused at an offset position far from the target chamber center due to its mechanical limitations; therefore, a spherical PM was used that allows the original focal pattern to be relayed at an offset position with respect to the target chamber center.
A spherical concave mirror (2-inch diameter and 202 mm curvature length) with a 1.053 $\mu$m anti-reflection coating on both surfaces was placed after the focus point, as shown in Fig.\,\ref{fig:experimental_setup}(a). The LFEX was focused at 3 mm above the target chamber center (offset position). The image at the offset position is relayed to the target chamber center with an image magnification of 1 by the spherical mirror. According to ray-trace code calculations, the deterioration of the image due to spherical aberration and astigmatism of the spherical mirror is negligible compared to the 70 $\mu$m diameter of the LFEX spot.
The laser energy fluence on the PM surface was optimized to be 90 J/cm$^2$ to obtain acceptable reflectivity (50$\%$) and spatial uniformity of the reflected pulse \cite{Morace2017}.

\subsection{Model for the hydrodynamics of the critical surface irradiated by multi-ps laser pulse}
We have previously derived the transition timing from the hole boring to
 the blowout phase $t_{s}$, under constant laser irradiation (Eq.\,(8) in Ref.\,\cite{iwata}). 
Here, the equation of transition timing is extended to a laser pulse with an arbitrary intensity temporal profile and the theoretical result is compared with that obtained by PIC simulation.

At the interface, plasma is pushed into a high-density region by the laser ponderomotive pressure caused by hole boring. 
The hole boring velocity is conventionally derived with assumption of the initial and terminal states of the plasma components (ions, bulk electrons) and REs \cite{Ping2012,Kemp2008,Bagnoud2017,Vincenti2014,Sentoku2006}.
It is assumed that the laser is reflected at the interface, which is moving with velocity $v_p$, and that electrons and ions are initially immobile. 
The flow velocity of REs is assumed to be $v_{h}$. 

In a frame moving with the interface at velocity $v_p$, ions and bulk electrons drift at $-v_p$ toward the interface, where they are reflected. $1-f_e$ is the fraction of electrons reflected elastically to the velocity $+v_p$, at the interface. The remaining fraction $f_e$ of electrons are heated by the laser and accelerated to relativistic velocities $p_h/\gamma m_e$. All ions are assumed to be reflected elastically to $+v_p$, which
 is the same assumption as that made by Vincenti \textit{et al.} \cite{Vincenti2014}.
The equations of the momentum flux conservation and the energy flux conservation are given by
$(1+R)I(t)/c\cos \theta \approx 2M_i n_i(t) v_p^2 + f_e n_e(t) v_h(t) p_h(t)$ and $(1-R)I(t)\cos \theta = f_e n_e(t) v_h(t) p_h(t) c$, respectively. 
Here, $R$ is the reflectivity of the incident laser on the plasma and $\theta$ is the laser incident angle. $M_i$ and $n_i$ ($m_e$ and $n_e$) are the ion (electron) mass and number density, respectively. 
Reflection and density steepening occur at the relativistic critical electron density $\gamma(t) n_c$ with $\gamma(t)=\sqrt{1+(1+R)a_0^2(t)/2}$, 
where $a_0$ is the normalized laser field amplitude. 
The initial electron density profile is assumed to be $n_e(x)=(\gamma n_c)\exp[(x-x_c(0))/l_s]$ with scale length $l_s$. 
Note that in Ref.\,\cite{iwata}, $\cos\theta =1$ and $f_{e}=1$ are assumed. 

The dashed lines in Fig.\,\ref{fig:Plasma_compression}(a) indicate the motion of the interface calculated by Eq.\,(\ref{eq: momentum conservation at the laser-plasma interaction surface_low}) with various reflectivities.
The red dashed line ($R$=0.7) reproduces the motion obtained by the PIC simulation until $t=$3.8\,ps.
This result shows that the velocity of the interface at each time can be determined by the momentum and energy balance among the laser, ions, and hot electrons, and that the integration of the interface velocity, Eq.\,\eqref{eq: momentum conservation at the laser-plasma interaction surface_low}, explains the motion of the interaction surface.

%\bibliographystyle{apsrev}
%\bibliography{library}

\end{document}